\documentstyle[aps,epsf,graphics]{revtex}  
\newcommand{\postscript}[2]
 {\setlength{\epsfxsize}{#2\hsize}
  \centerline{\epsfbox{#1}}}

\begin{document}        

\baselineskip 14pt
\title{Search for Scalar Top}
\author{Christopher Holck}
\address{University of Pennsylvania\\
        for the \\
        CDF Collaboration}
%
\maketitle              

\begin{abstract}        
We report results of three searches for scalar top quark.  Two of the
searches look for direct production of scalar top quark followed by
the decay of the scalar quark to charm quark and neutralino or bottom
and chargino.  The third search looks for top quark decaying to scalar
top and neutralino followed by the decay of scalar top to bottom quark
and neutralino.  We find no evidence for the presence of scalar top
quark in any of the searches.  Therefore, depending on the search we
set limits on the production cross-section, ${\mathcal BR}(t
\rightarrow \tilde{t}_1 + \tilde{\chi}_{1}^{0})$, or $m_{\tilde{t}}$
vs.\ $m_{\tilde{\chi}_{1}^{0}}$.
\end{abstract}   	

\section{Introduction}               
\label{sec:intro}
Supersymmetry (SUSY) assigns to every fermionic SM particle a bosonic
superpartner and vice versa~\cite{mssm}. Therefore, the SM quark
helicity states $q_L$ and $q_R$ acquire scalar SUSY partners
$\tilde{q_L}$ and $\tilde{q_R}$ which are also the mass eigenstates
for the first two generations to a good approximation. However, a
large mixing can occur in the third generation leading to a large
splitting between the mass eigenstates~\cite{lstop}.  This can lead to
a scalar top quark ($\tilde{t}_1$) which is not only the lightest
scalar quark but also lighter than the top quark.

At the Tevatron, scalar top quark is directly produced in pairs via
$gg$ and $q\overline{q}$ fusion diagrams.  The cross section depends
only on $m_{\tilde{t}}$ at leading order~\cite{Beenakker:1997ut}. The
dominant next-to-leading order SUSY corrections depend on the other
scalar quark masses and are small ($\sim 1\%$)\@.  For
$m_{\tilde{t}_1}$= 110 GeV$/c^2$, $\sigma_{\tilde{t}_1
{\overline{\tilde{t}}}_1}$ = 7.4 pb in the next-to-leading order.  If
$m_{\tilde{t}_1} + m_{\tilde{\chi}_{1}^{0}}< m_t$, $\tilde{t}_1$ can
be indirectly produced via the decay $t \rightarrow \tilde{t}_1 +
\tilde{\chi}_{1}^{0}$.

Whenever kinematically allowed, $\tilde{t}_1 \rightarrow b
\tilde{\chi}_{1}^{+}$.  If this channel is closed but scalar neutrino
($\tilde{\nu}$) is light, then $\tilde{t}_1 \rightarrow b l
\tilde{\nu} $ dominates.  If neither of these channels is allowed,
then scalar top decays via a one-loop diagram to charm and a
neutralino: $\tilde{t}_1 \rightarrow c \tilde{\chi}_{1}^{0}$.  We
assume that $\tilde{\chi}_{1}^{0}$ is the lightest SUSY particle and
that R-parity is conserved.  Hence the $\tilde{\chi}_{1}^{0}$ is
undetected and causes an imbalance of energy.

At CDF, we have performed three separate analyses: (I) direct
production of $\tilde{t}_1 {\overline{\tilde{t}}}_1$ with
$BR(\tilde{t}_1 \rightarrow c + \tilde{\chi}_{1}^{0}) = 100\%$, (II)
$t \rightarrow \tilde{t}_1 \tilde{\chi}_{1}^{0}$ with $\tilde{t}_1
\rightarrow b \tilde{\chi}_{1}^{+}$, and (III) direct production of
$\tilde{t}_1 {\overline{\tilde{t}}}_1$ with $BR(\tilde{t}_1
\rightarrow b + \tilde{\chi}_{1}^{+}) = 100\%$.

\section{D\lowercase{irect search for $\tilde{t}_1 \rightarrow c \tilde{\chi}_{1}^{0}$}}
\label{sec:stop_clsp}

The signature for $\tilde{t}_1$ pair production if $\tilde{t}_1
\rightarrow c \tilde{\chi}_{1}^{0}$ is 2 acolinear charm jets,
significant missing transverse energy ($\not\!\!{E}_T$), and no
high--p$_T$ lepton(s)\@.  For this analysis, we have searched data
corresponding to a total integrated luminosity of 88$\pm$3.6 pb$^{-1}$
collected using the CDF detector during the 1994--95 Tevatron run.
CDF is a general purpose detector consisting of tracking, vertexing,
calorimeter components and a muon system~\cite{Abe:1988me}.  Events
from this analysis were collected using a trigger which required
$\not\!\!{E}_T > 35$~GeV.

We select events with 2 or 3 jets which have $E_T \geq 15$~GeV and
$|\eta| \leq 2$ and no other jets with $E_T > 7$~GeV and $|\eta| <
3.6$.  The $\not\!\!{E}_T$\ cut is increased beyond trigger threshold
to 40~GeV and we require that $\not\!\!{E}_T$ is neither parallel nor
anti-parallel to any of the jets in the event to reduce the
contribution from the processes where missing energy comes from jet
energy mismeasurement: $min\ \Delta\phi(\not\!\!{E}_T,j) > 45^\circ$,
$\Delta\phi(\not\!\!{E}_T,j_1) < 165^\circ$, and
$45^\circ<\Delta\phi(j_1,j_2)<165^\circ$.  The jet indices are ordered
by decreasing $E_T$.  We reject events with an identified electron or
muon.

To select events with a charm jet, we determine the probability that
the ensemble of tracks within a jet is consistent with coming from the
primary vertex.  We require that at least one jet has a probability of
less than 5\%.

The largest source of background for this analysis is $W/Z+jets$
production where the vector boson decays to a lepton (e/$\mu$) that is
not identified or to a $\tau$ lepton which decays hadronically.  There
is also a small contribution from Q$CD\ multi$--$jet$ production.

We observe 11 events which is consistent with $14.5\pm4.2$ events from
Standard Model processes.  We interpret this as an excluded region in
the $m_{\tilde{t}_1}$--$m_{\tilde{\chi}_{1}^{0}}$ parameter space as
shown in Fig.~\ref{fig:stop_clsp}.  The maximum $m_{\tilde{t}_1}$
excluded is 119~GeV/$c^2$ for $m_{\tilde{\chi}_{1}^{0}} =
40$~GeV/$c^2$.

\begin{tabular}{cc}

\begin{minipage}{0.45\textwidth}
\begin{table}
 \begin{tabular}{||l|l||}
  Sample  & $N_{exp}$ \\ \hline
  $W^{\pm}(\rightarrow e^{\pm}\nu_{e})+\geq 2$ jets& $0.3\pm0.3\pm0.1$\\ \hline
  $W^{\pm}(\rightarrow \mu^{\pm}\nu_{\mu})+\geq 2$ jets & $0.9\pm0.5\pm0.3$\\ \hline
  $W^{\pm}(\rightarrow \tau^{\pm}\nu_{\tau})+\geq 1$ jets  & $7.6\pm1.6\pm2.2$ \\ \hline
  $Z^0(\rightarrow \nu\overline{\nu})+\geq 2$ jets  & $1.2\pm0.4\pm0.4$ \\ \hline
  $t\overline{t}$ & $0.7\pm0.2\pm0.4$ \\ \hline
  Diboson (WW+WZ+ZZ) & $0.4\pm0.1\pm0.1$ \\ \hline
  Total W/Z/Top bkg & $11.1\pm1.8\pm3.3$ \\ \hline
  Total QCD bkg & $3.4\pm1.7$ \\ \hline
  Total Expected & $14.5\pm4.2$ \\ \hline \hline
  Total Observed & $11$ \\
 \end{tabular}
\caption{The expected composition of the data sample after \emph{all} cuts have been applied.}
\label{tbl:top_clsp}
\end{table}
\end{minipage}

\hspace{0.05\textwidth}

\begin{minipage}{0.45\textwidth}
\begin{figure}[t]	
\postscript{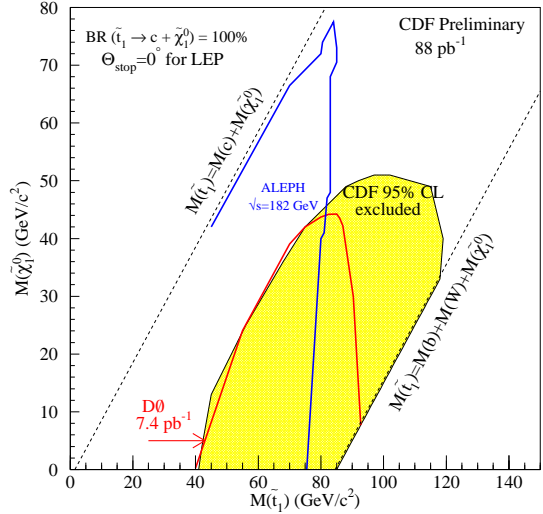}{1.}   
\caption[]{95\% CL limit for $\tilde{t}_1 \rightarrow c \tilde{\chi}_{1}^{0}$ search.  The D\O\ limits are from~\cite{Abachi:1996jp} and the ALEPH limits are from~\cite{Barate:1998zn}.}
\label{fig:stop_clsp}
\end{figure}
\end{minipage}
\end{tabular}

\section{S\lowercase{earch for $t \rightarrow \tilde{t}_1 \tilde{\chi}_{1}^{0}$}}
\label{sec:top_stop}

If $BR(t \rightarrow \tilde{t}_1 \tilde{\chi}_{1}^{0}) \neq 0$, then
top pair production can have three types of events: (1) $t\overline{t}
\rightarrow bW^+ \, \overline{b}W^-$ (SM--SM) (2) $t\overline{t}
\rightarrow \tilde{t}_1 \tilde{\chi}_{1}^{0} \, \overline{b}W^-$
(SM--SUSY) (3) $t\overline{t} \rightarrow \tilde{t}_1
\tilde{\chi}_{1}^{0} \, \overline{\tilde{t}}_1 \tilde{\chi}_{1}^{0}$
(SUSY--SUSY)\@.  Further, if $\tilde{t}_1 \rightarrow b
\tilde{\chi}_{1}^{+}$, then all three cases can lead to a signal
topology of one isolated, high--p$_T$ lepton (e/$\mu$), three or more
high $E_T$ jets (one of which is a b jet) and missing transverse
energy ($\not\!\!{E}_T$)\@.  This analysis uses $109.4\pm7.2$
pb$^{-1}$ of inclusive, high--p$_T$ lepton data collected during the
1992--1993 and 1994--1995 Tevatron runs.  The cuts used in this
analysis are based on the cuts described
in~\cite{Abe:1995eh},~\cite{Abe:1995hr}.

To select events, we start by requiring~$\not\!\!{E}_T > 25$~GeV and a
single electron (muon) with $E_T (P_T) > 20$~GeV~(GeV/c)\@.  We reject
$Z+jets$ background by looking for an additional lepton of the same
flavor but opposite charge and require that the dilepton invariant
mass be greater than 105~GeV/c$^2$ or less than 75~GeV/c$^2$.  We also
demand that the transverse mass ($M_T$) formed by the lepton and the
$\not\!\!{E}_T$ be greater than 40~GeV/c$^2$.  The $M_T$ cut rejects
events where the lepton does not come from the decay of a W (such as
Drell-Yan events).

We demand that there be at least three jets with $|\eta| < 2$ and the
following $E_T$ requirements (jets are ordered by decreasing $E_T$):
$E_T$(jet 1) $> 20$~GeV, $E_T$(jet 2) $> 20$~GeV , $E_T$(jet 3) $>
15$~GeV\@.  We cut on the cosine of the angle between the jet and the
proton beam as computed in the rest frame of the event ($\equiv
\cos{\theta^*_i}$)\@.  Ordering the three highest $E_T$ jets by
$|\cos{\theta^*_i}|$ we demand that $|\cos{\theta^*_i}|_1 < 0.9,\
|\cos{\theta^*_j}|_2 < 0.8,\ |\cos{\theta^*_k}|_3 < 0.7$.

To further reduce the $W+jets$ background, we require that the three
highest $E_T$ jets are well separated: $\Delta R(j_i,j_k) \geq 0.9\
(i,k=1,2,3)$.  We also require the transverse momentum of the W
(p$_T(W)$), which is constructed from the lepton p$_T$ and
$\not\!\!{E}_T$, to be large: p$_T(W) > 50$~GeV/c.  We take advantage
of the presence of extra $\tilde{\chi}_{1}^{0}$s in SUSY top events by
increasing the $\not\!\!{E}_T$ cut: $\not\!\!{E}_T > 45$~GeV\@.
Finally, we require at least one $b$--jet by demanding at least one
SVX--tagged jet~\cite{Abe:1995hr}.

After applying all these selection requirements, our data sample
should (as shown from Monte Carlo studies) be composed almost entirely
of top events (with both SM and SUSY top decays)\@.  For the purposes
of setting a limit, we assume that the $W+jets$ background is zero.

To distinguish (SM--SM) events from (SUSY--SM)+(SUSY--SUSY) we exploit
the difference in the $E_T$(jet 2) and $E_T$(jet 3) distributions for
these two classes of events.  The $E_T$(jet 2)/$E_T$(jet 3)
distributions will be softer for SUSY events relative to SM events.
We define a \emph{Relative Likelihood} ($\equiv \mathcal{R_L}$):
\begin{eqnarray}
  {\mathcal{R_L}} & = & \frac{{\mathcal{L}}_{Abs}^{t \rightarrow
\tilde{t}_1 \tilde{\chi}^+}}{{\mathcal{L}}_{Abs}^{t \rightarrow b W^+}}\\
{\mathcal{L}}_{Abs} & = & (\frac{1}{\sigma}\frac{d\sigma}{dE_T(2)}) \times (\frac{1}{\sigma}\frac{d\sigma}{dE_T(3)})\ \ \ ({\mathcal{L}}_{Abs}\ {\mathrm is\ derived\ from\ Monte\ Carlo}.)
\end{eqnarray}
SM events will have $-\ln{{\mathcal{R_L}}} > 0$ (\emph{SM--like
region}) and SUSY events will have $-\ln{{\mathcal{R_L}}} < 0$
(\emph{SUSY--like region}).  This is shown in
Fig.~\ref{fig:top_stop_like}.

To set a limit on ${\mathcal BR}(t \rightarrow \tilde{t}_1
\tilde{\chi}_{1}^{0})$ as a function of
$m_{\tilde{t}_1}$,$m_{\chi_1^{\pm}}$,$m_{\chi_1^0}$ we employ the
following method.  We generate SM top Monte Carlo and SUSY top Monte
Carlo for a given ${\mathcal
BR}$,$m_{\tilde{t}_1}$,$m_{\chi_1^{\pm}}$,$m_{\chi_1^0}$.  We
normalize the sum of SM top and SUSY top in the \emph{SM--like region}
to what is observed in data (9 events)\@.  Using this normalization,
we can determine the number of SUSY top plus SM top events we expect
in the \emph{SUSY--like region} ($\equiv \mu({\mathcal BR})$)\@.  We
determine the Poisson probability (including errors) that we observe 0
events in the \emph{SUSY--like region} when we expect $\mu({\mathcal
BR})$ events and set a 95\% Confidence Level (C.L.)
~\cite{Caso:1998tx} on the ${\mathcal BR}$ for a given
$m_{\tilde{t}_1}$,$m_{\chi_1^{\pm}}$,$m_{\chi_1^0}$.  The 95\% C.L. on
${\mathcal BR}$ as a function of $m_{\tilde{t}_1}$,$m_{\chi_1^{\pm}}$
for $m_{\chi_1^0} = 40$~GeV/c$^2$ is shown if Fig.~\ref{fig:top_stop}.

\begin{tabular}{cc}

\begin{minipage}{0.45\textwidth}
\begin{figure}[t]	
\postscript{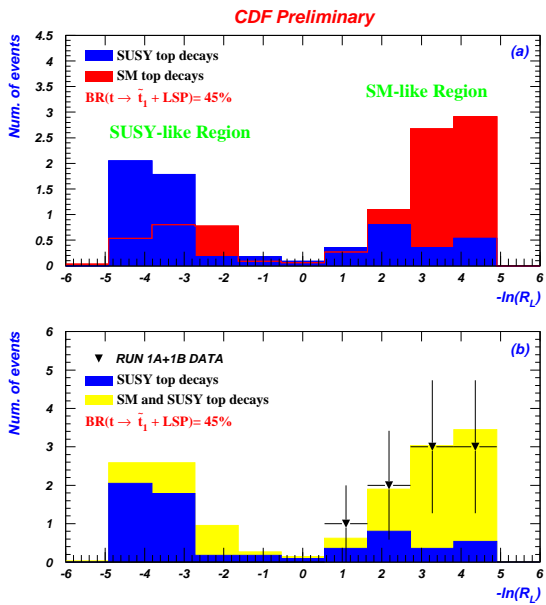}{1.}   
\caption[]{The expected $-\log {\mathcal R_L}$ distribution for top events after all cuts are applied is shown in (a).  ${\mathcal BR}$($t \rightarrow \tilde{t}_1 \chi_1^0$) = 45\%.  In (b) we overlay the data distribution.}
\label{fig:top_stop_like}
\end{figure}
\end{minipage}

\hspace{0.05\textwidth}

\begin{minipage}{0.45\textwidth}
\begin{figure}[t]	
\postscript{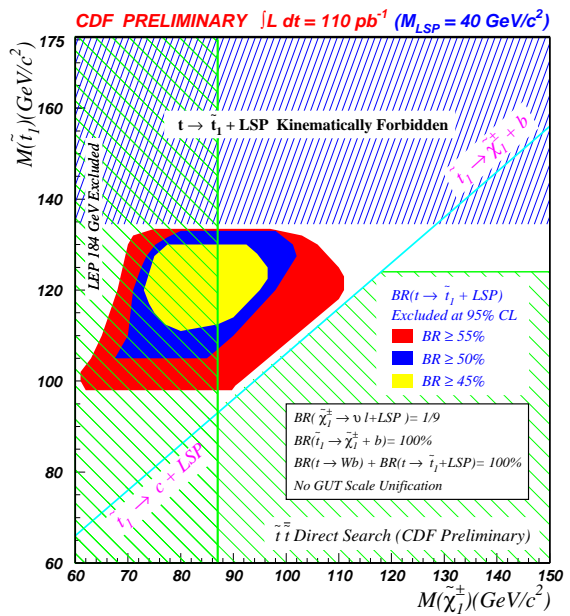}{1.}   
\caption[]{The 95\% C.L. exclusion contour for $m_{\chi_1^0} = 40$~GeV/c$^2$ and ${\mathcal BR} \geq 45,50,55\%$.}
\label{fig:top_stop}
\end{figure}
\end{minipage}
\end{tabular}

\section{D\lowercase{irect search for $\tilde{t}_1 \rightarrow b \tilde{\chi}_{1}^{+}$}}
\label{sec:stop_bchar}

If $\tilde{t}_1 \rightarrow b \tilde{\chi}_{1}^{+}$, then $\tilde{t}_1
\overline{\tilde{t}}_1 \rightarrow b\overline{b} \tilde{\chi}_{1}^{+}
\tilde{\chi}_{1}^{-}$.  We assume that $\tilde{\chi}_{1}^{+}$ is
\emph{gaugino--like} and has the same couplings as the Standard Model
W.  If one $\chi$ decays leptonically ($\chi^+_1 \rightarrow l \nu
\chi^0_1,\ l=e\mu$) and the other $\chi$ decays hadronically
($\chi^+_1 q q' \chi^0_1$) then the final signal topology is one
high--p$_T$ lepton (e/$\mu$ only), three or more high--E$_T$ jets (one
of which is a $b$--jet), and $\not\!\!{E}_T$.  We search 90.1$\pm$5.9
pb$^{-1}$ of data collected during the 1994--1995 Tevatron run.

To select events, we require one electron (muon) with $E_T$ ($P_T$)
$>20$ GeV (GeV/c) and $\not\!\!{E}_T > 20$ GeV\@.  We also cut on the
jet multiplicity.  We demand one jet with $E_T > 15$ GeV, $|\eta| <
2.$, one jet with $E_T > 8$ GeV, $|\eta| < 2.4$ and $\leq 4$ jets with
$E_T > 8$ GeV\@.  After these cuts, we are left with 2249 electron
events and 1754 muon events.  To make our final sample, we require at
least one jet with a SVX tag as defined in Sec.~\ref{sec:top_stop}.
Our final sample consists of 47 electron events and 41 muons events.

The data consist of three components: $W+jets$, $t\overline{t}$, and
$\tilde{t}_1\overline{\tilde{t}}_1$.  To set a limit on the number of
$\tilde{t}_1$ events in this sample, we perform an unbinned likelihood
fit using two uncorrelated kinematic variables: $M_T$ and
$\Delta\phi(lepton,j_2)$ ($j_2$ is the second highest $E_T$ jet in the
event)\@.  The likelihood function ($ \equiv {\mathcal L}$) is:
\begin{eqnarray}
{\mathcal L}(N_W,N_T,N_S) & = & (\prod_i \frac{N_W W(i) + N_T T(i) + N_S S(i)}{N_W + N_T + N_S}) \cdot \exp {(-(N-N_{obs})^2/2 N_{obs})} \cdot \exp {(-(N_T-\overline{N_T})^2/2 \Delta\overline{N_T}^2)}
\end{eqnarray}

The product runs over the number of observed events ($N_{obs}$).
$N_W/N_T/N_S$ are the \emph{fitted} number of
$W+jets$/$t\overline{t}$/$\tilde{t}_1 \overline{\tilde{t}}_1$ events
and $N = N_W+N_T+N_S$.  $W/T/S$ are the joint probability densities
for the $M_T$ and $\Delta\phi(lepton,j_2)$ distributions.  Since these
variables are uncorrelated, the joint probability density is equal to
the product of the individual probability densities.  For top and
$\tilde{t}_1$, the probability densities are taken from Monte Carlo
after all selection criteria (including SVX tagging) have been
applied.  For $W+jets$, we use the distributions from data
\emph{before} tagging is applied.  The first exponential term
constrains, within errors, the total number of events to $N_{obs}$.
The second exponential constrains, within errors, the top contribution
to what we expect using the theoretical cross--section (5.4$\pm$0.5
pb$^{-1}$~\cite{Berger:1995vm}).

To set a limit on a given $\tilde{t}_1$ mass (for a fixed
$m_{\chi^{\pm}_1}$,$m_{\chi^0_1}$), we minimize the negative log
likelihood.  The minimization returns $N_S$ plus its error.  Using the
total acceptance from Monte Carlo, we convert this into an excluded
cross--section.  In Fig.~\ref{fig:stop_mt}, we show the $M_T$
distribution. The $W+jets$, top, and $\tilde{t}_1$ ($m_{\tilde{t}_1} =
90$ GeV/c$^2$) distributions are normalized to the number of events
returned by the $- \log {\mathcal L}$ minimization.  In
Fig.~\ref{fig:stop_bchar} we show the cross-section excluded by data
versus the LO cross-section for two pairs of
$m_{\chi^{\pm}_1}$,$m_{\chi^0_1}$.

\begin{tabular}{cc}

\begin{minipage}{0.45\textwidth}
\begin{figure}
\postscript{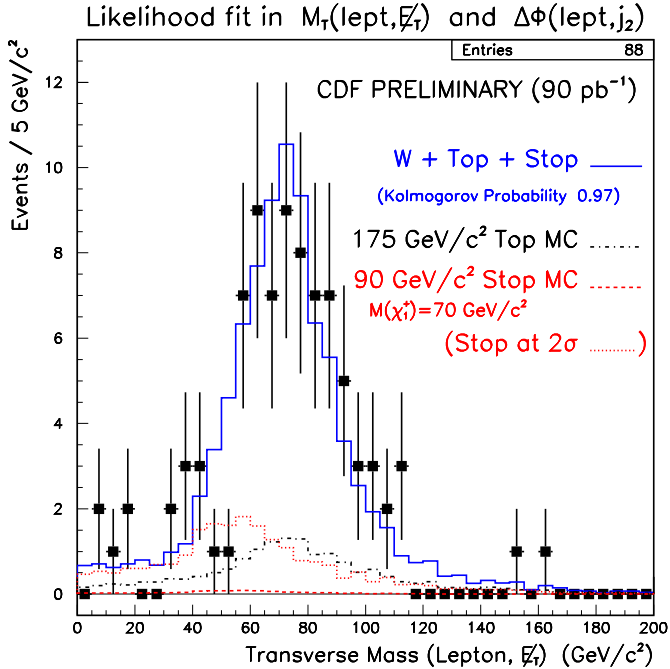}{1.}
\caption[]{$M_T$ distribution after tagging.  The normalization for the $W+jets$, $t$, and $\tilde{t}_1$ distributions is described in the text.}
\label{fig:stop_mt}
\end{figure}
\end{minipage}

\hspace{0.05\textwidth}

\begin{minipage}{0.45\textwidth}
\begin{figure}	
\postscript{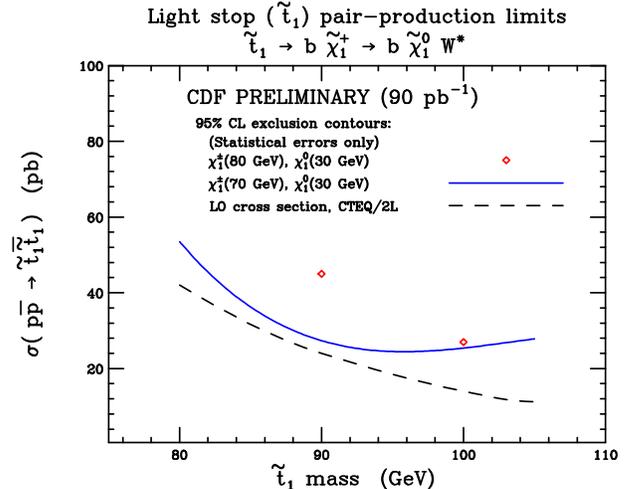}{1.}
\caption[]{The 95\% C.L. excluded cross--section as a function of $m_{\tilde{t}_1}$ for ($m_{\chi_1^{\pm}},m_{\chi_1^0})$ = (70 GeV/c$^2$,\, 30 GeV/c$^2$) and (80 GeV/c$^2$,\, 30 GeV/c$^2$).}
\label{fig:stop_bchar}
\end{figure}
\end{minipage}

\end{tabular}

\section{Aknowledgements}
We would like to thank Carmine Pagliarone and Michael Gold for their
help in understand these analyses.  We also thank the Fermilab staff
and technical staffs of the participating institutions for their
contributions.

\end{document}